\documentclass[]{mn2e}
\usepackage{epsfig}

\title[M\,33 monitoring. I]{The UK Infrared Telescope M\,33 monitoring
project. I. Variable red giant stars in the central square kiloparsec}
\author[Javadi, van Loon \& Mirtorabi]{Atefeh Javadi$^{1,2}$\thanks{E-mail:
atefeh@astro.keele.ac.uk}, Jacco Th.\ van Loon$^{2}$ and Mohammad Taghi
Mirtorabi$^{1}$\\
$^{1}$Physics Department, Alzahra University, Vanak, Tehran, Iran\\
$^{2}$Astrophysics Group, Lennard-Jones Laboratories, Keele University,
     Staffordshire ST5 5BG, UK}
\date{Resubmitted 8 September 2010}
\pagerange{\pageref{firstpage}--\pageref{lastpage}}
\pubyear{2010}
\begin{document}
\maketitle
\label{firstpage}
\begin{abstract}
We have conducted a near-infrared monitoring campaign at the UK InfraRed
Telescope (UKIRT), of the Local Group spiral galaxy M\,33 (Triangulum). The
main aim was to identify stars in the very final stage of their evolution, and
for which the luminosity is more directly related to the birth mass than the
more numerous less-evolved giant stars that continue to increase in
luminosity. The most extensive dataset was obtained in the K-band with the
UIST instrument for the central $4^\prime\times4^\prime$ (1 kpc$^2$) -- this
contains the nuclear star cluster and inner disc. These data, taken during the
period 2003--2007, were complemented by J- and H-band images. Photometry was
obtained for 18,398 stars in this region; of these, 812 stars were found to be
variable, most of which are Asymptotic Giant Branch (AGB) stars. Our data were
matched to optical catalogues of variable stars and carbon stars, and to
mid-infrared photometry from the {\it Spitzer} Space Telescope. In this first
of a series of papers, we present the methodology of the variability survey
and the photometric catalogue -- which is made publicly available at the
Centre de Donn\'ees astronomiques de Strasbourg (CDS) -- and discuss the
properties of the variable stars. Most dusty AGB stars had not been previously
identified in optical variability surveys, and our survey is also more
complete for these types of stars than the {\it Spitzer} survey.
\end{abstract}

\section{Introduction}

Messier 33 is one of three stereotypical spiral galaxies that inhabit the
Local Group. Located in the constellation of Triangulum, it spans about a
degree on the sky. Its favourable inclination angle of $56^\circ$ makes M\,33
a prime subject for the study of the detailed structure and stellar content of
a spiral galaxy like our own.

The distance to M\,33 has been determined {\it via} several methods: Freedman,
Wilson \& Madore (1991) determined a distance of 850 kpc by using Cepheid-type
variables, i.e.\ a distance modulus $\mu=24.65$ mag, but Scowcroft et al.\
(2009) revisited this to $\mu=24.53\pm0.11$ mag. By using RR\,Lyrae-type
variables, Sarajedini et al.\ (2000) found $\mu=24.84$ mag. Other techniques
that have been employed used the tip of the Red Giant Branch (RGB; Rizzi et
al.\ 2007; Galleti, Bellazzini \& Ferraro 2004; McConnachie et al.\ 2004;
Tiede, Sarajedini \& Barker 2004; Kim et al.\ 2002), Long-Period Variables
(LPVs; Pierce, Jurcevi\'c \& Crabtree 2000), and detached eclipsing binaries
(Bonanos et al.\ 2006) -- the latter, a largely geometric technique, yielded
$\mu=24.92\pm0.12$ mag. U et al.\ (2009) reviewed these techniques, which
result in differences in distance modulus of $\Delta\mu\sim0.6$ mag,
corresponding to an uncertainty in the distance of $\approx30$\%. By using
blue (massive) supergiants, U et al.\ themselves determined $\mu=24.93\pm0.11$
mag. In this paper we shall adopt $\mu=24.9$ mag.

Tracing stellar populations of a wide range in ages, from as recently formed
as 30 Myr ago to as ancient as 10 Gyr, Asymptotic Giant Branch (AGB) stars
(Marigo et al.\ 2008) are exquisite probes of the star formation history of
galaxies. With their high luminosity ($\approx1000$--60,000 L$_\odot$) and low
temperature ($T\sim3000$--4500 K), AGB stars dominate the appearance of
galaxies at near-infrared (near-IR) wavelengths. This is aided further by the
low extinction at IR wavelengths compared to that at optical wavelengths. More
massive stars, up to $\sim30$ M$_\odot$, become red supergiants (RSGs;
Levesque et al.\ 2005; Levesque 2010) and they can be used to trace the more
recent star formation history over about 10--30 Myr ago.

Stars at this advanced level of evolution exhibit strong radial pulsations on
timescales of typically 150--1500 days (e.g., Wood et al.\ 1992; Wood 1998;
Pierce et al.\ 2000; Whitelock et al.\ 2003). As a result of this pulsation
AGB stars lose up to 80\% of their mass to the interstellar medium (ISM; Bowen
1988; Bowen \& Willson 1991; Vassiliadis \& Wood 1993; van Loon et al.\ 1999,
2005), making AGB stars important contributors to chemical enrichment of
galaxies; RSGs contribute less in total, but their mass loss sets the
conditions within which the ensuing supernova develops (van Loon 2010). The
most-evolved AGB stars pulsate in the fundamental mode, which has the largest
amplitude, with less-evolved AGB stars and RGB stars pulsating in an overtone
with a smaller amplitude (Wood 1999). Though no less powerful, the amplitude
of pulsation expressed in terms of magnitude, being a relative scale,
diminishes with increasing time-averaged luminosity (Wood et al.\ 1992; van
Loon et al.\ 2008) making it more difficult to detect the pulsation of RSGs
than of lower-luminosity AGB stars (though the photometric errors for RSGs are
also smaller).

Detecting \emph{variable} AGB stars is a powerful tool in reconstructing the
star formation history of a galaxy as these stars are in the final stages of
their evolution and hence their luminosity is more directly related to their
birth mass than that of less-evolved AGB stars that still undergo significant
evolution in luminosity.

Various variability surveys have been conducted in M\,33, mostly at optical
wavelengths and with cadences too short to adequately identify LPVs. Among the
most comprehensive monitoring campaigns count those of Hartman et al.\ (2006),
the DIRECT project (Macri et al.\ 2001; Mochejska et al.\ 2001a,b), and
Sarajedini et al.\ (2006) who identified 64 RR\,Lyrae variables. McQuinn et
al.\ (2007), on the other hand, identified variables in M\,33 on the basis of
five epochs of mid-IR observations performed with the {\it Spitzer} Space
Telescope.

The main objectives of the project are: to construct the mass function of LPVs
and derive from this the star formation history in M\,33; to correlate spatial
distributions of the LPVs of different mass with galactic structures
(spheroid, disc and spiral arm components); to measure the rate at which dust
is produced and fed into the ISM; to establish correlations between the dust
production rate, luminosity, and amplitude of an LPV; and to compare the {\it
in situ} dust replenishment with the amount of pre-existing dust. This is
Paper I in the series, presenting the methodology used in the monitoring
campaign and search for variable stars, the properties of the photometric
catalogue of stars in the inner square kpc, and cross-identifications in
various other relevant photometric and variability catalogues. Subsequent
papers in the series will discuss the galactic structure and star formation
history in the inner square kpc (Paper II), the mass-loss mechanism and dust
production rate (Paper III), and the extension to a nearly square degree area
covering much of the M\,33 optical disc (Paper IV).

\section{Observations}

Observations were made with three of UKIRT's imagers: UIST, UFTI, and WFCAM.
The WFCAM observations cover a much larger part of M\,33 and are discussed in
Paper IV in this series.

\subsection{UIST}

%
%
\begin{table}
\caption[]{Log of UIST observations of each of 4 quadrants (``Q'').}
\begin{tabular}{cccccc}
\hline\hline
Date (y\,m\,d)      &
Q                   &
Filter              &
Epoch               &
$t_{\rm int}$ (min) &
Airmass             \\
\hline
2003 10 06 & 1 & K &         1 & 21 & 1.49--1.35 \\
2003 10 06 & 1 & J &         1 & 15 & 1.11--1.08 \\
2003 10 06 & 2 & K &         1 & 21 & 1.24--1.16 \\
2003 10 06 & 2 & J &         1 & 15 & 1.08--1.05 \\
2003 11 18 & 3 & K &         1 & 21 & 1.21--1.30 \\
2004 01 06 & 3 & K &         2 & 21 & 1.02--1.02 \\
2004 01 06 & 3 & J &         1 & 15 & 1.05--1.07 \\
2004 01 06 & 4 & K &         1 & 21 & 1.02--1.03 \\
2004 01 06 & 4 & J &         1 & 15 & 1.08--1.11 \\
2004 07 25 & 1 & K &         2 & 27 & 1.22--1.13 \\
2004 07 25 & 2 & K &         2 & 27 & 1.12--1.07 \\
2004 07 25 & 3 & K &         3 & 27 & 1.05--1.02 \\
2004 07 26 & 4 & K &         2 & 27 & 1.05--1.02 \\
2004 09 09 & 1 & K &         3 & 27 & 1.34--1.21 \\
2004 09 09 & 2 & K &         3 & 27 & 1.21--1.12 \\
2004 09 09 & 3 & K &         4 & 27 & 1.09--1.04 \\
2004 09 09 & 4 & K &         3 & 27 & 1.04--1.02 \\
2004 12 24 & 1 & K &         4 & 27 & 1.03--1.07 \\
2004 12 24 & 2 & K &         4 & 27 & 1.07--1.13 \\
2005 01 08 & 1 & J &         2 & 21 & 1.35--1.49 \\
2005 01 08 & 2 & J &         2 & 21 & 1.51--1.71 \\
2005 01 18 & 1 & H &         1 & 15 & 1.03--1.05 \\
2005 01 18 & 2 & H &         1 & 15 & 1.05--1.07 \\
2005 01 25 & 3 & K &         5 & 27 & 1.05--1.10 \\
2005 01 25 & 4 & K &         4 & 27 & 1.11--1.18 \\
2005 02 22 & 3 & K &         6 & 27 & 1.40--1.62 \\
2005 02 22 & 4 & K &         5 & 27 & 1.63--1.98 \\
2005 07 04 & 1 & K &         5 & 27 & 1.63--1.41 \\
2005 07 04 & 2 & K &         5 & 27 & 1.40--1.26 \\
2005 07 05 & 3 & K &         7 & 27 & 1.70--1.46 \\
2005 07 05 & 4 & K &         6 & 27 & 1.45--1.29 \\
2005 07 27 & 1 & K &         6 & 27 & 1.25--1.15 \\
2005 07 27 & 2 & K &         6 & 27 & 1.15--1.08 \\
2006 08 08 & 1 & K &         7 & 27 & 1.27--1.17 \\
2006 08 08 & 1 & J &         3 & 27 & 1.07--1.03 \\
2006 08 08 & 2 & K &         7 & 27 & 1.16--1.09 \\
2006 08 08 & 2 & J &         3 & 27 & 1.03--1.02 \\
2006 08 09 & 1 & H &         2 & 27 & 1.40--1.26 \\
2006 08 09 & 2 & H &         2 & 27 & 1.25--1.15 \\
2006 08 09 & 3 & K &         8 & 27 & 1.12--1.06 \\
2006 08 09 & 3 & H &         1 & 27 & 1.02--1.02 \\
2006 08 09 & 4 & K &         7 & 27 & 1.06--1.03 \\
2006 08 09 & 4 & H &         1 & 27 & 1.02--1.03 \\
2006 08 10 & 3 & J &         2 & 27 & 1.17--1.10 \\
2006 08 10 & 4 & J &         2 & 27 & 1.10--1.05 \\
2006 09 06 & 3 & K &         9 & 22 & 1.42--1.25 \\
2006 09 06 & 4 & K &         8 & 22 & 1.25--1.14 \\
2006 09 10 & 1 & K &         8 & 22 & 1.04--1.02 \\
2006 09 10 & 2 & K &         8 & 21 & 1.02--1.03 \\
2006 10 03 & 3 & K & \llap{1}0 & 21 & 1.67--1.48 \\
2006 10 03 & 4 & K &         9 & 21 & 1.48--1.34 \\
2006 10 04 & 1 & K &         9 & 21 & 1.66--1.48 \\
2006 10 04 & 2 & K &         9 & 21 & 1.47--1.34 \\
2006 10 23 & 1 & K & \llap{1}0 & 21 & 1.14--1.09 \\
2006 10 23 & 2 & K & \llap{1}0 & 21 & 1.09--1.05 \\
2006 10 23 & 3 & K & \llap{1}1 & 21 & 1.06--1.09 \\
2006 10 23 & 4 & K & \llap{1}0 & 21 & 1.10--1.15 \\
2007 06 18 & 3 & K & \llap{1}2 & 27 & 1.53--1.35 \\
2007 07 03 & 4 & K & \llap{1}1 & 27 & 1.31--1.20 \\
2007 07 13 & 2 & K & \llap{1}1 & 27 & 1.11--1.06 \\
2007 07 25 & 1 & K & \llap{1}1 & 27 & 1.05--1.02 \\
\hline
\end{tabular}
\end{table}

The monitoring campaign comprises observations with the UIST imager, of each
of four quadrants with slight overlap keeping the unresolved nucleus of M\,33
in one of the corners of the field. The approximate centres (slight variations
occur due to the absolute pointing accuracy of UKIRT) are respectively
($1^{\rm h}33^{\rm m}47.^{\rm s}0$, $+30^\circ38^\prime47^{\prime\prime}$),
($1^{\rm h}33^{\rm m}54.^{\rm s}8$, $+30^\circ38^\prime47^{\prime\prime}$),
($1^{\rm h}33^{\rm m}54.^{\rm s}8$, $+30^\circ40^\prime27^{\prime\prime}$),
($1^{\rm h}33^{\rm m}47.^{\rm s}0$, $+30^\circ40^\prime27^{\prime\prime}$),
for fields 1, 2, 3 and 4. Observations were made in the K-band (UKIRT filter
K98) over the period October 2003 -- July 2007. All quadrants were imaged 11
times except the $3^{\rm rd}$ quadrant which was imaged on a $12^{\rm th}$
occasion. Due to the (small) overlap between these quadrants some stars will
have been measured more often, usually (but not always) on the same night.
Each quadrant was at least once imaged in the J- and H-bands (UKIRT filters
J98 and H98, respectively) to provide colour information. A log is given in
Table 1.

Each image has $1024\times1024$ pixels of $0.12^{\prime\prime}$, and the
combined mosaic covers approximately $4^\prime\times4^\prime$ -- a square kpc
at the distance of M\,33. A 9-point dither (small offset) pattern was employed
to be able to map the background intensity by median-combination of the
individual exposures. With exposure times of 20 s, and 5--9 cycle repeats, the
total integration time per epoch varied between 15 and 27 min. The frames were
combined and corrected for background light and for spatial variations of the
system response and throughput (``flatfield'') using the {\sc orac-dr}
software and the {\sc bright\_point\_source} recipe. Photometric standard
stars were observed on several nights, some of which were of photometric
quality. For the standard star images only the central $512\times512$ pixels
were saved, and these were combined using the {\sc jitter\_self\_flat\_aph}
recipe. The seeing constraint on all observations was for it to be
$<0.8^{\prime\prime}$, but in some images the stars appear distorted
(elongated) due to movements of the camera during integrations.

\subsection{UFTI}

%
%
\begin{table}
\caption[]{Log of UFTI observations of each of 9 fields (``X'').}
\begin{tabular}{cccccc}
\hline\hline
Date (y\,m\,d)      &
X                   &
Filter              &
Epoch               &
$t_{\rm int}$ (min) &
Airmass             \\
\hline
2005 08 11 & 1 & K & 1\rlap{a} & 9 & 1.39--1.33 \\
2005 08 11 & 2 & K & 1         & 9 & 1.32--1.27 \\
2005 08 11 & 3 & K & 1         & 9 & 1.26--1.22 \\
2005 08 11 & 4 & K & 1         & 9 & 1.22--1.18 \\
2005 08 11 & 5 & K & 1         & 9 & 1.18--1.14 \\
2005 08 11 & 6 & K & 1         & 9 & 1.14--1.11 \\
2005 08 11 & 7 & K & 1         & 9 & 1.11--1.09 \\
2005 08 11 & 8 & K & 1         & 9 & 1.08--1.07 \\
2005 08 11 & 9 & K & 1         & 9 & 1.06--1.05 \\
2005 08 11 & 1 & K & 1\rlap{b} & 9 & 1.05--1.04 \\
2005 08 12 & 1 & K & 2\rlap{a} & 9 & 1.58--1.50 \\
2005 08 12 & 2 & K & 2         & 9 & 1.49--1.41 \\
2005 08 12 & 3 & K & 2         & 9 & 1.40--1.34 \\
2005 08 12 & 4 & K & 2         & 9 & 1.33--1.28 \\
2005 08 12 & 5 & K & 2         & 9 & 1.27--1.23 \\
2005 08 12 & 6 & K & 2         & 9 & 1.23--1.19 \\
2005 08 12 & 7 & K & 2         & 9 & 1.18--1.15 \\
2005 08 12 & 8 & K & 2         & 9 & 1.15--1.12 \\
2005 08 12 & 9 & K & 2         & 9 & 1.11--1.09 \\
2005 08 12 & 1 & K & 2\rlap{b} & 9 & 1.09--1.07 \\
2005 08 13 & 1 & K & 3\rlap{a} & 9 & 1.06--1.05 \\
2005 08 13 & 2 & K & 3         & 9 & 1.04--1.03 \\
2005 08 13 & 3 & K & 3         & 9 & 1.03--1.02 \\
2005 08 13 & 4 & K & 3         & 9 & 1.02--1.02 \\
2005 08 13 & 5 & K & 3         & 9 & 1.02--1.02 \\
2005 08 13 & 6 & K & 3         & 9 & 1.02--1.02 \\
2005 08 13 & 7 & K & 3         & 9 & 1.02--1.02 \\
2005 08 13 & 8 & K & 3         & 9 & 1.03--1.03 \\
2005 08 13 & 9 & K & 3         & 9 & 1.03--1.04 \\
2005 08 13 & 1 & K & 3\rlap{b} & 9 & 1.05--1.06 \\
\hline
\end{tabular}
\end{table}

On three consecutive nights in August 2005 the UFTI imager was used instead,
in the K-band (UKIRT filter K98) only. The UFTI camera provides
$1024\times1024$ pixels of $0.09^{\prime\prime}$, and a slightly larger area
than covered with UIST was mapped with nine slightly overlapping fields. The
mosaic was started and finished with the central field, which thus received
twice the integration time on a particular night. The same 9-point dither
pattern and 20 s integration time per frame were employed, but due to the
slower survey speed the integration time per field amounted to only 9 min per
night. Two repeats on the following nights resulted in 27 min total
integration time per field (54 min for the central field). In regions of
overlap some stars received longer total integration times. A standard star
was observed in the same manner as with UIST, and the same {\sc orac-dr}
recipes were used as before.

All UFTI observations were combined to create one mosaic. In doing this, a
``super-sky'' was constructed by combining the peripheral 8 fields in the
pixel domain (i.e.\ without aligning them in the celestial coordinate system)
via their modal pixel values rejecting the 4 highest (out of 24) values, and
applying a box modal filter to smooth out small-scale residuals due to
repeated incidences of stars. This super-sky was then subtracted from each
field, their background intensity levels were brought in relative agreement by
subtracting the mode from each field (10 counts per pixel were added to the
central field to establish agreement with the peripheral fields), and finally
the fields were aligned in the celestial coordinate system and combined
normalised by the total exposure time per pixel.

%
%
\begin{figure*}
\centerline{\psfig{figure=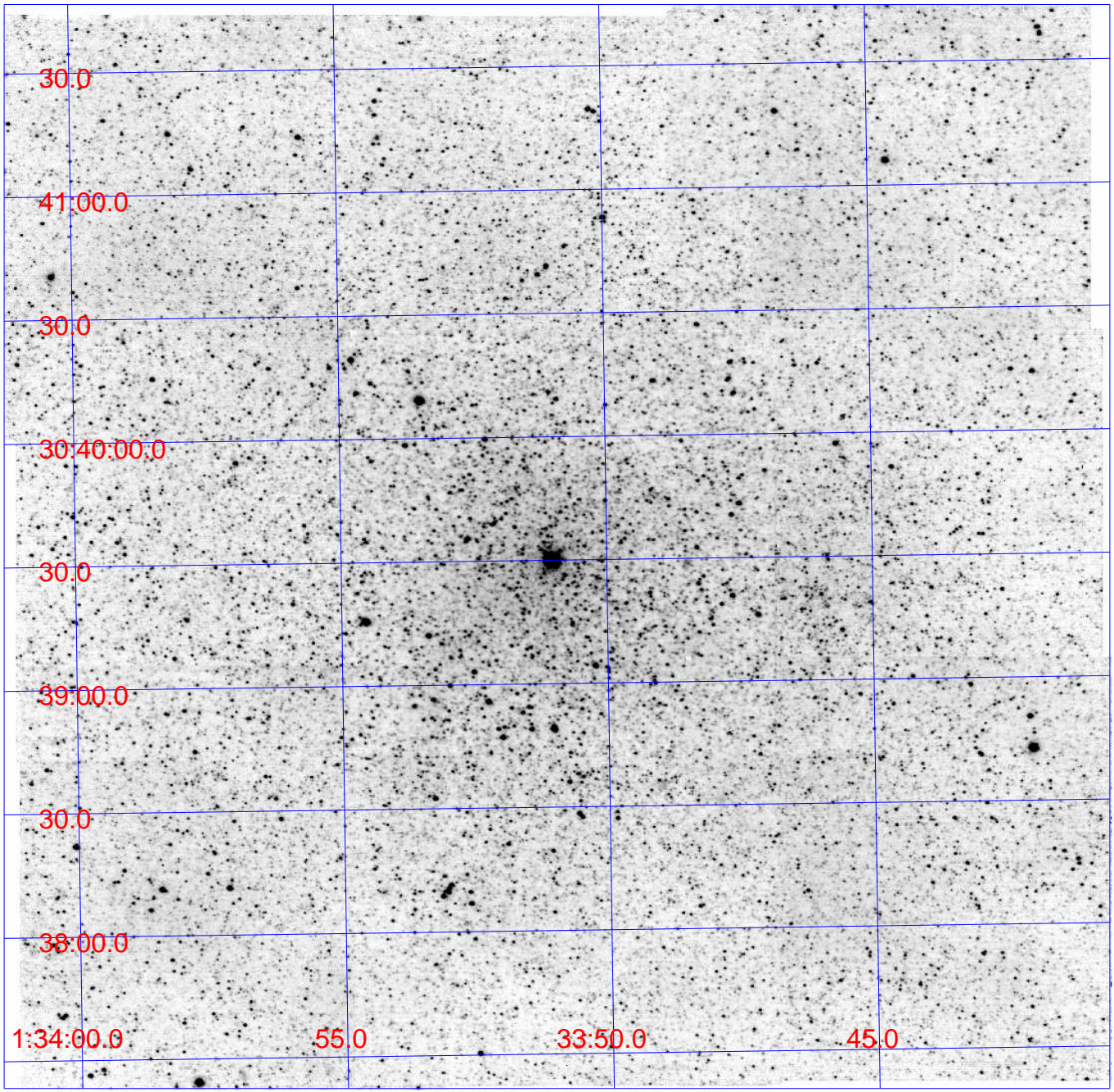,width=180mm}}
\caption[]{UFTI K-band mosaic of the central square kiloparsec of M\,33.}
\end{figure*}

The resulting UFTI K-band mosaic is shown in Fig.\ 1. The central
$5^{\prime\prime}$ diameter region of the nuclear star cluster is unresolved.
This is surrounded by a region of $\sim1^\prime$ diameter with a considerably
higher stellar density, the ``bulge''. Outside that, the image is dominated by
the inner part of the disc of M\,33, although it is difficult to make out the
spiral arm pattern that is a conspicuous feature of M\,33 at larger distances
from the nucleus.

\section{Photometry}

Photometry was obtained for all stars within each frame by automated fitting
of a model of the Point Spread Function (PSF), using the {\sc daophot/allstar}
software suite (Stetson 1987). Depending on the conditions we used either a
constant or quadratically-varying PSF. For the images with a constant-PSF
model we selected 15--30 isolated PSF stars; for the other images we selected
50--100 PSF stars. The final PSF model was made from the frame from which all
neighbours of the PSF stars had been subtracted. All PSF stars were selected
based on their $\chi$ (goodness-of-fit) and ``sharpness'' value; ideally,
$\chi=1$ and sharpness $=0$ for stars with realistic uncertainty estimates.
The resulting PSF-subtracted image was then examined, rejecting any PSF stars
that had not subtracted perfectly from the image.

The individual images were aligned using the {\sc daomaster} routine, which
computes the astrometric transformation equation coefficients from the {\sc
daophot/allstar} results. We combined the individual images using the {\sc
montage2} routine (Stetson 1994), and then co-added these three J-, H- and
K-band images with the {\sc imarith} task in {\sc Iraf} to create a master
mosaic of the central $4^\prime\times4^\prime$ of M\,33. A master catalogue of
stars was then created by application of {\sc daophot/allstar} to this master
mosaic, using over 100 PSF stars to create the (quadratically-varying) PSF
model. We thus obtained photometry for 18,518 stars.

The master catalogue was used as input for {\sc allframe}, which
simultaneously performed PSF-fitting photometry on these stars within each of
the individual images. The PSF created from the master mosaic did not fit
perfectly to the stars in the individual images, mainly because of variations
in the Full Width at Half Maximum (FWHM). Therefore, we used instead the PSFs
created for each image separately.

%
%
\begin{figure}
\centerline{\psfig{figure=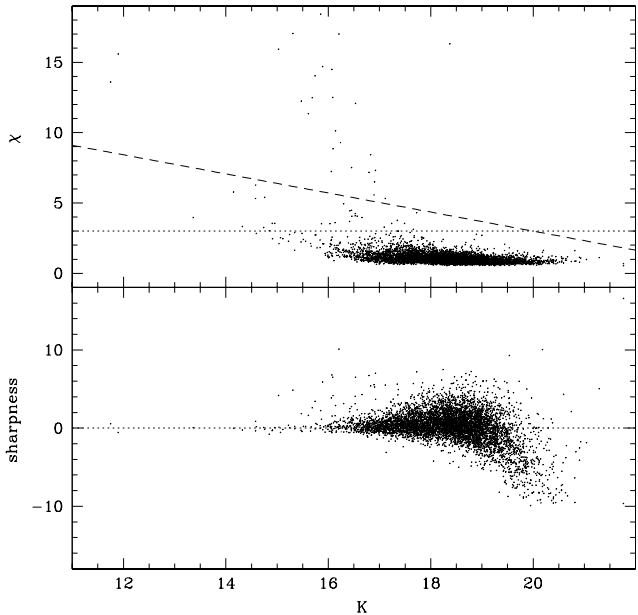,width=84mm}}
\caption[]{The $\chi$ (top) and sharpness (bottom) values, {\it vs.}\ K-band
magnitude for stars in one of the individual frames. The dashed line shows the
selection limit on the $\chi$ value.}
\end{figure}

%
%
\begin{figure}
\centerline{\psfig{figure=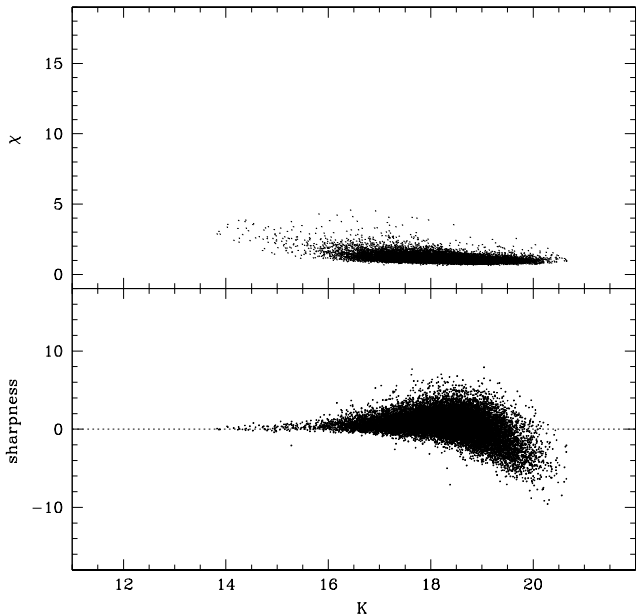,width=84mm}}
\caption[]{Average $\chi$ (top) and sharpness (bottom) values, {\it vs.}\
K-band magnitude, after applying a selection on $\chi$ (see text).}
\end{figure}

In Fig.\ 2 we show the $\chi$ and sharpness values {\it vs}.\ K-band magnitude
for one of our individual frames. We removed non-stellar objects by setting
limits on the $\chi$ value, per individual frame. We did not set limits on the
sharpness value. Stars were selected with $\chi$ values below the dashed line
in Fig.\ 2 (but this differs for other frames). For this frame, most stellar
objects have $\chi<3$, but one can notice some bright objects with slightly
larger $\chi$ values; this may be due to inaccuracies in the adopted PSF
model. We also removed objects from the catalogue if they obeyed the $\chi$
criterion in only 10\% of the measurements for that object. In this way our
final catalogue has 18,398 stellar objects. The average $\chi$ and sharpness
values of objects in the catalogue are shown in Fig.\ 3.

Aperture corrections to the PSF-fitting photometry were determined using the
{\sc daogrow} routine (Stetson 1990) to construct growth-curves for each frame
from which all stars had been subtracted except the PSF stars. We then applied
the {\sc collect} routine (Stetson 1993) to calculate the "aperture
correction", i.e.\ the difference between the PSF-fitting and large-aperture
magnitude of these stars. We added this value to each of the PSF-fitting
magnitudes.

Photometric calibration was then performed in a two-step manner. First, the
standard star measurements were used to calibrate the frames on those nights.
For the frames without standard star measurements we adopted the documented
zero points\footnote{http://www.jach.hawaii.edu/UKIRT/astronomy/calib/
phot\_cal/cam\_zp.html} (although this is not really needed).
Airmass-dependent atmospheric extinction corrections were applied adopting the
extinction coefficients derived by Krisciunas et al.\ (1987). In the second
step, we calibrated the frames relative to one another. For this, we selected
approximately 1000 stars in common between all frames belonging to a specific
quadrant, within the magnitude interval $K\in [16...18]$, and averaged the
magnitudes of these selected stars in each frame. The photometry was brought
in line with each other by applying corrections between $-0.052$ to 0.067 mag
(but usually much smaller). Checks on the overlap regions showed that no
additional corrections had to be applied to the photometry between the four
quadrants. Finally, the photometry was transformed onto the 2MASS system by
using the tranformation equations derived by Carpenter et al.\ (2001).

\subsection{Survey completeness}

%
%
\begin{figure}
\centerline{\psfig{figure=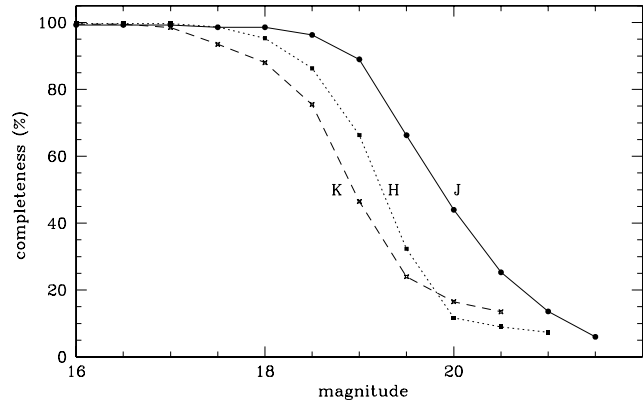,width=84mm}}
\caption[]{Completeness as a function of J- (dots, solid line), H- (squares,
dotted) and K-band (crosses, dashed) magnitude.}
\end{figure}

To estimate the completeness of our catalogue and the effect of blending we
added 300 artificial stars in each of 5 trials to the master mosaic using the
{\sc daophot/addstar} task (Stetson 1987). We added stars in 0.5-mag bins
starting from $K=16$ mag until $K=20.5$ mag (until 21 mag in $H$ and 21.5 mag
in $J$). Stars were positioned randomly in the image and Poisson noise was
added. Then, we repeated the {\sc daofind/allstar/allframe} procedure on the
new frame as described before. Once the photometry was done and the final list
created we used {\sc daomaster} to evaluate what fraction of stars was
recovered. As one can see in Fig.\ 4 our catalogue is essentially complete
down to $K\sim17$ mag, complete to $>88$\% down to $K\sim18$ mag (near the RGB
tip, see below), dropping to below 50\% at $K=19$ mag. The J-band reaches
similar completeness levels but at about a magnitude fainter, and the H-band
lies somewhere in between those two.

%
%
\begin{figure}
\centerline{\psfig{figure=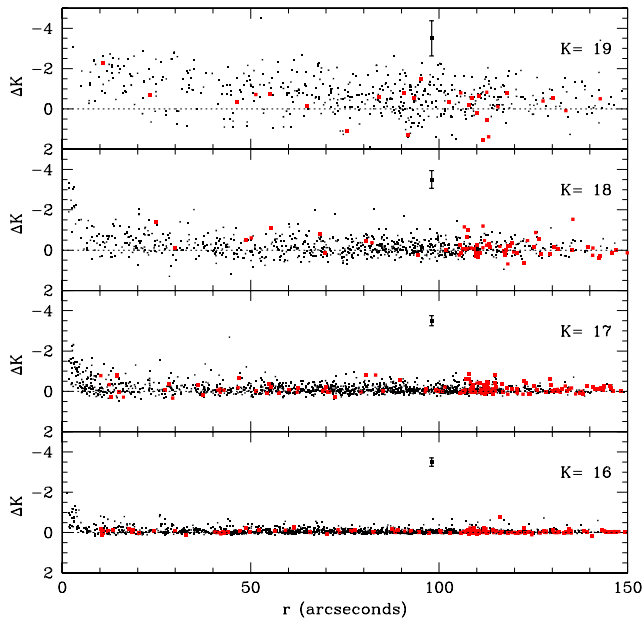,width=84mm}}
\caption[]{Difference between input magnitude and recovered magnitude (from a
K-band image) {\it vs.}\ distance from the centre of M\,33, for four values of
the input magnitude. Stars observed near the edges of the frames are
highlighted in red, and the dispersion is indicated in the form of an errorbar
in each panel.}
\end{figure}

We also examined the accuracy of our photometry by adding artificial stars to
one of the individual frames. Like before, we repeated the photometry for this
frame after adding 300 stars in each of 5 trials, where the added stars were
given magnitudes $K\in (16,17,18,19)$. The difference between input magnitude
and recovered magnitude is small, $|\Delta K|<0.2$ mag, except very near the
centre of M\,33, $r<5^{\prime\prime}$, where it reaches $\Delta K<-1$ mag
(Fig.\ 5), i.e.\ the recovered stars are brighter than the input stars. We
attribute this to severe blending in the unresolved nucleus of M\,33. The same
pattern is seen for $K=17$ and 18 mag just with increased values for $\Delta
K$. Stars that are located near the edges of the images are recovered with the
same degree of flux conservation as stars elsewhere in the images, so we
conclude that the photometry is accurate also for stars near the frame edges.
At $K=19$ mag most of the recovered stars are brighter than the input stars
irrespective of distance from the centre. By then, crowding has become
substantial: the stellar density reaches $\sim0.06$--0.11 star per square
arcsec down to $K=19.5$ mag.

The scientific objectives of our project concentrate on the use of AGB stars
and RSGs, that have $K<18.3$ mag, hence our analysis is not seriously
compromised by the completeness limit and blending of stars except within the
inner few arcseconds of M\,33. It is also important to note that, while {\it
absolute} photometry in crowded regions may be compromised, {\it relative}
photometry between images can be more accurate, and variability throughout a
{\it series} of measurements may be detected with confidence at a level -- or
below that -- of the errors on individual measurements.

\section{Variability analysis}

Variable star candidates were identified with the {\sc newtrial} routine
(Stetson 1993) which uses the technique developped by Welch \& Stetson (1993)
and Stetson (1996). This method first calculates the $J$ index:
\begin{equation}
J=\frac{\Sigma_{k=1}^n w_k\ {\rm sign}(P_k)\sqrt{|P_k|}}{\Sigma_{k=1}^n w_k}.
\end{equation}
Here, observations $i$ and $j$ have been paired and each pair $k$ has been
given a weight $w_k$; the product of the normalised residuals,
$P_k=(\delta_i\delta_j)_k$, where $\delta_i=(m_i-\langle
m\rangle)/\epsilon_i)$ is the deviation of measurement $i$ from the mean,
normalised by the error on the measurement, $\epsilon_i$. Note that $\delta_i$
and $\delta_j$ may refer to measurements taken in different
filters.\footnote{Following Stetson (1996), $P_k=\delta^2-1$ if $i=j$.} The
$J$ index has a large positive value for variable stars and tends to zero for
data containing random noise only.

In circumstances where we are dealing with a small number of observations or
corrupt data we gain from also calculating the Kurtosis index:
\begin{equation}
K=\frac{\frac{1}{N} \Sigma_{i=1}^n |\delta_i|}{\sqrt{\frac{1}{N}\Sigma_{i=1}^N
\delta_i^2}}.
\end{equation}
The value of $K$ depends on the shape of the light-curve: $K=0.9$ for a
sinusoidal light variation, where the source spends most time near the
extrema, $K=0.798$ for a Gaussian distribution, which is concentrated towards
the average brightness level (as would random noise), and $K\rightarrow0$ for
data affected by a single outlier (when $N\rightarrow\infty$).

The variability index that we calculate in this paper depends on both the $J$
and $K$ indices and is defined by (Stetson 1996):
\begin{equation}
L=\frac{J\times K}{0.798}.
\end{equation}

%
%
\begin{figure}
\centerline{\psfig{figure=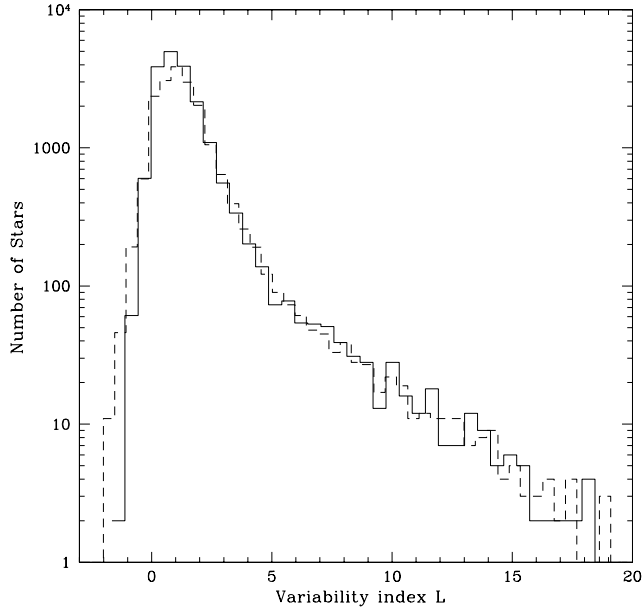,width=84mm}}
\caption[]{Histogram of variability index for two ways of pairing three
measurements $a$, $b$ and $c$: (dashed:) $ab$, $bc$; (solid:) $ab$, $bc$,
$ac$. The latter is the extended method proposed by Stetson (1996) and which
is adopted here.}
\end{figure}

Measurements can be paired if they are taken close in time compared to the
(expected) period of variability -- which for the type of stars we search for
within the context of this programme is of order 100 days or longer. If within
a pair of observations only one measurement is available for a particular star
then the weight of the pair for that star is set to 0.5. Stetson (1996)
extended this principle to involve more than two measurements at a time: three
measurements $a$, $b$ and $c$ taken within a time-span less than the shortest
period can be paired as $ab$, $bc$, and $ac$ rather than just $ab$ and $bc$.
There is no discernable difference in the variability index for different
pairing schemes (Fig.\ 6), and we decided to use Stetson's extended method of
pairing here.

%
%
\begin{figure}
\centerline{\psfig{figure=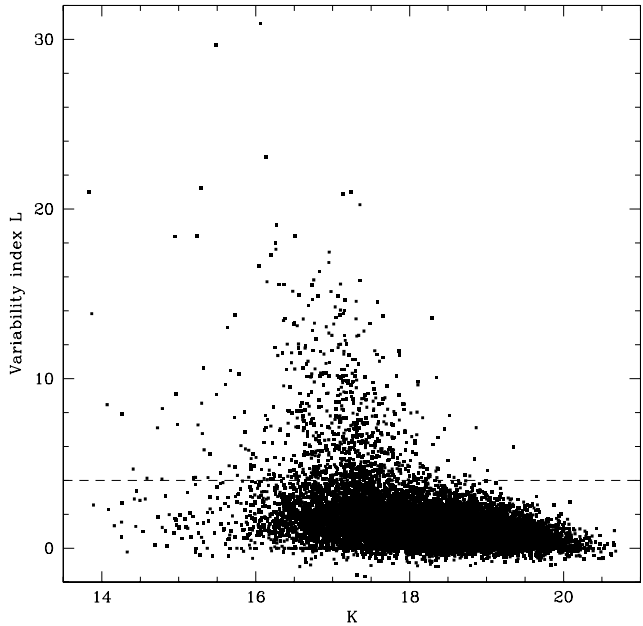,width=84mm}}
\caption[]{Variability index $L$ {\it vs.}\ K-band magnitude. The dashed line
indicates our threshold for identifying variable stars, at $L>4$.}
\end{figure}

%
%
\begin{figure}
\centerline{\psfig{figure=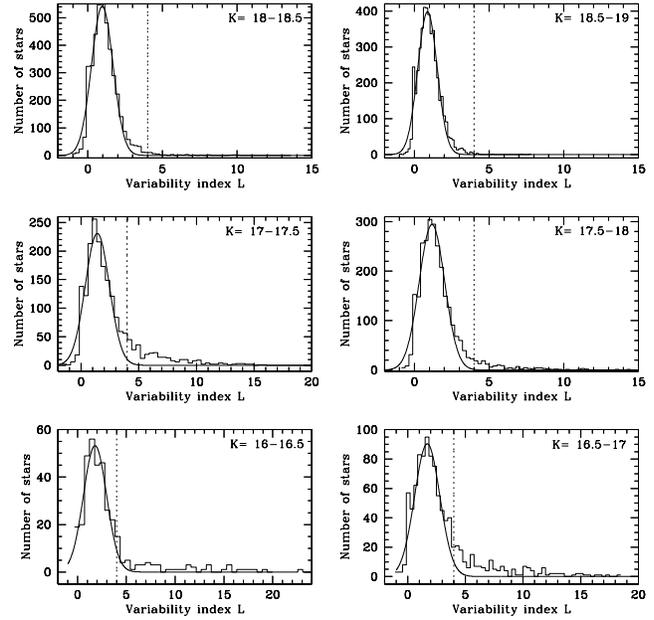,width=84mm}}
\caption[]{Histograms of the variability index $L$, for several K-band
magnitude bins. The solid lines show Gaussian functions fitted to the
histograms.}
\end{figure}

Fig.\ 7 shows the variability index $L$ {\it vs.}\ K-band magnitude, where the
dashed line indicates our threshold for detected variability: $L>4$. There is
a remarkable ``branch'' of variable stars between $K\sim16$--18 mag which are
likely AGB stars that show Mira-type variability, with hardly any fainter
variables and a modest number of brighter variables. Histograms of the
variability index for several K-band magnitude intervals in the range 16--18.5
mag are shown in Fig.\ 8. To determine the optimal variability threshold, a
Gaussian function was fitted to each of these histograms. While the Gaussian
function is a near-perfect fit to the symmetrical distributions at low values
for $L$, the distributions show a pronounced tail towards higher values for
$L$. The departure from the Gaussian shape occurs typically around
$L\approx4$.

%
%
\begin{figure}
\centerline{\psfig{figure=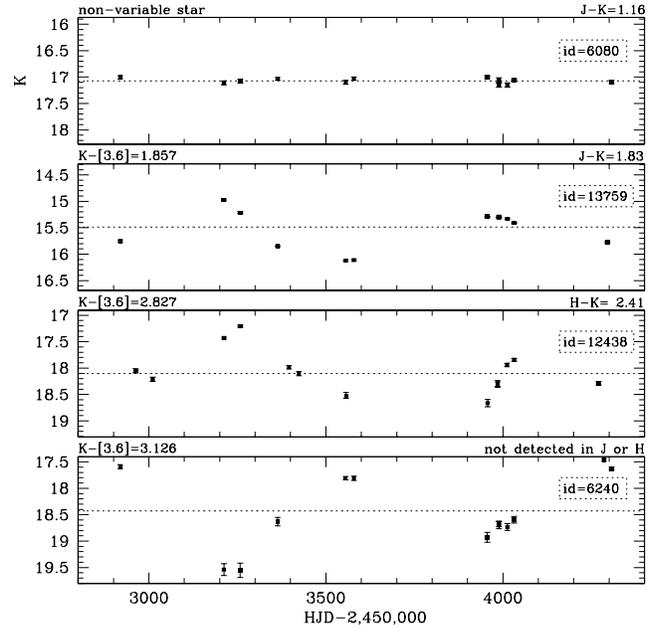,width=84mm}}
\caption[]{Example light-curves of three very red variable stars with large
amplitude and probably long period (around 700 days), with a non-variable star
in the top panel for comparison.}
\end{figure}

We thus identified 812 variable stars in the central square kiloparsec of
M\,33. (A further 6 were rejected on the basis of their large $\chi$ values.).
Three examples of large-amplitude, likely long-period variability are shown in
Fig.\ 9, with a non-variable star for comparison. These variable stars have a
period around 700 days and much redder colours than the bulk of the
non-variable stars.

\subsection{Amplitudes of variability}

A measure for the ampltidue of variability can be obtained by assuming a
sinusoidal light-curve shape. The standard deviation of the unit sine function
is 0.701. Hence, for a standard deviation in our data, $\sigma$, the amplitude
estimate would be $A=2\times\sigma/0.701$.\footnote{We follow the custom to
define the amplitude as the difference between the minimum and maximum
brightness.}

%
%
\begin{figure}
\centerline{\psfig{figure=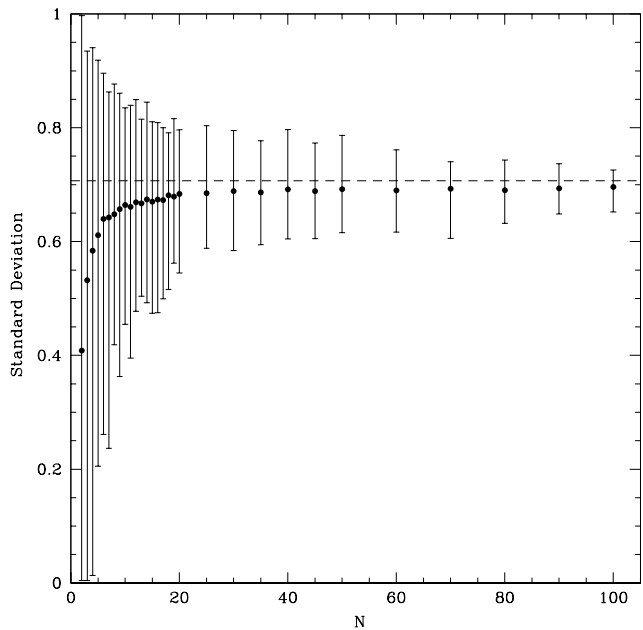,width=84mm}}
\caption[]{Standard deviation {\it vs.}\ number of measurements that randomly
sample a unit sine curve.}
\end{figure}

In practice, however, for small sets of data the standard deviation depends on
the number of measurements. Fig.\ 10 shows the result of simulations: it is
clear that when only a handful of measurements are available, the standard
deviation that is calculated will, on average, drop below the asymptotic value
(0.701 for the sine function) that is reached for well-sampled light-curves,
as well as becoming less reliable. We set $N=6$ as a minimum for applying our
method to estimate the amplitude from the standard deviation. From Fig.\ 10
one can see that, by then, the standard deviation will have converged to
within 10\% of the asymptotic value.

%
%
\begin{figure}
\centerline{\psfig{figure=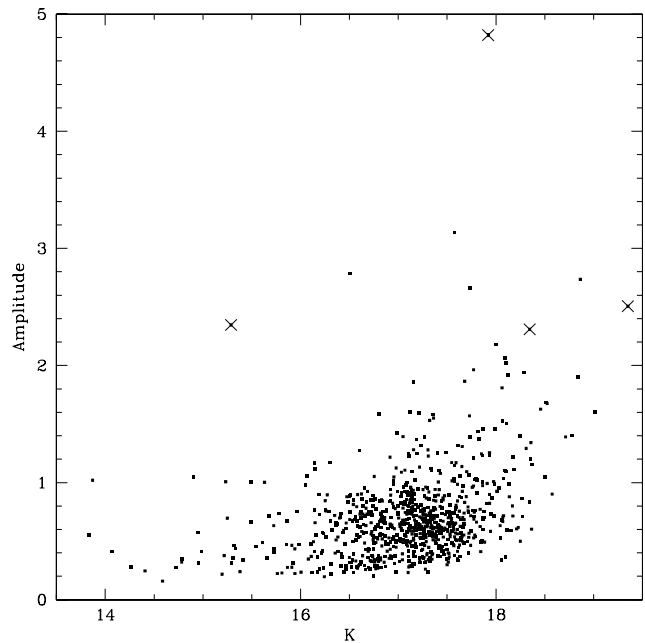,width=84mm}}
\caption[]{Estimated amplitude, $A_{\rm K}$, of variability {\it vs.}\ K-band
magnitude. Crosses indicate variables with $A_{\rm K}>2$ mag but with only six
or fewer measurements.}
\end{figure}

The estimated K-band amplitude of variability is plotted {\it vs.}\ K-band
magnitude in Fig.\ 11. Variability could have been detected for $A_{\rm
K}>0.2$ mag. There is a clear tendency for the amplitude to diminish with
increasing brightness, which is a known (Wood et al.\ 1992; Wood 1998;
Whitelock et al.\ 2003) and to some extent understood (van Loon et al.\ 2008)
trend. Among the variables with $A_{\rm K}>2$ mag, four have six or fewer
measurements and their amplitudes are therefore unreliable -- these are
highlighted in Fig.\ 11 with crosses, and includes the one with the highest
estimate for the amplitude. Disregarding those, the amplitudes stay below
$A_{\rm K}\sim3$ mag and generally $A_{\rm K}<2$ mag. Very dusty AGB stars are
known to reach such large amplitudes (Wood et al.\ 1992; Wood 1998; Whitelock
et al.\ 2003), but they are very rare. Light-curves of a few large-amplitude
variables in our survey were displayed in Fig.\ 9.

%
%
\begin{figure}
\centerline{\psfig{figure=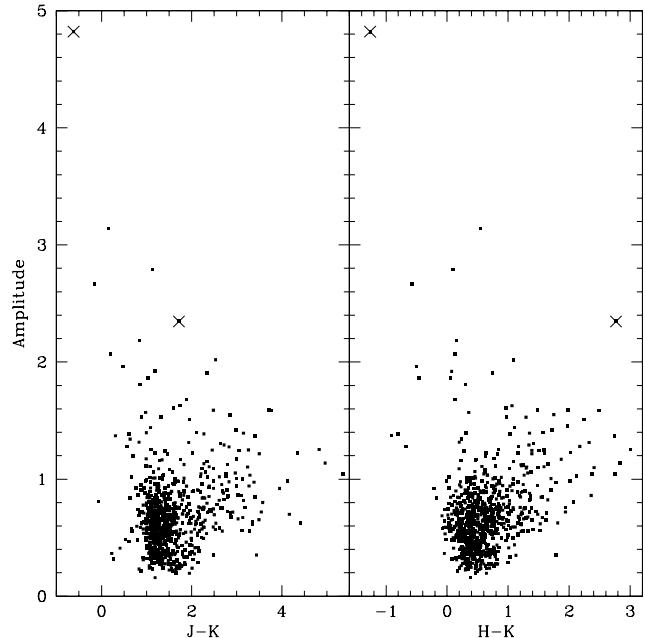,width=84mm}}
\caption[]{Estimated amplitude, $A_{\rm K}$, of variability {\it vs.}\ colour.
Crosses indicate variables with $A_{\rm K}>2$ mag but with only six or fewer
measurements (two further such stars have no J- or H-band detection).}
\end{figure}

The estimated K-band amplitude of variability is plotted {\it vs.}\ the $J-K$
and $H-K$ colour in Fig.\ 12. There is a clear sequence branching off the bulk
of stars towards redder colours and associated larger amplitudes (cf.\ Fig.\
9). This is not surprising as large-amplitude variability is known to be
associated with profuse dust formation (Wood et al.\ 1992; Wood 1998;
Whitelock et al.\ 2003), and the less-luminous larger-amplitude stars will be
rendered obscured much more readily than their more luminous siblings (van
Loon et al.\ 1997).

A small number of very large amplitude stars form a vertical sequence around
$J-K\sim1$ and $H-K\sim0$ mag (Fig.\ 12), but these are suspect. We inspected
all of the stars with $A_{\rm K}>1.7$ mag by eye, and found that they fall
into either of two categories: (1) nine bright stars, $15.3<K<18$ mag, that
are affected by blending. In some frames these were mistaken for another faint
star, resulting in a spuriously large amplitude. The small number of variable
stars that are affected by such complications ($\sim1$\% ) indicates that
these are rare incidences; (2) nine faint stars, $K>18$ mag, which were not
visibly affected by blending. Two of these in actual fact do have red colours,
$J-K=2.3$ mag ($A_{\rm K}=1.9$ mag) and $J-K=2.5$ mag ($A_{\rm K}=2.0$ mag),
and they may be dusty large-amplitude variables. The same could explain the
non-detection in the J- and H-band of four other stars in this category. That
leaves three stars that appear blue for their large amplitude. These are
located near the edge of the frame, resulting in particularly poor photometry
on one or two occasions. Again, these respresent very few occurrences of such
effects.

\section{Description of the catalogue}

%
%
\begin{table}
\caption[]{Description of the photometric catalogue.}
\begin{tabular}{rl}
\hline\hline
Column No. & Descriptor \\
\hline
\multicolumn{2}{l}{\it Part I: stellar mean properties (18,398 lines)} \\
 1         & Star number \\
 2         & Right Ascension (J2000) \\
 3         & Declination (J2000) \\
 4         & Mean J-band magnitude \\
 5         & Error in $\langle J\rangle$ \\
 6         & Mean H-band magnitude \\
 7         & Error in $\langle H\rangle$ \\
 8         & Mean K-band magnitude \\
 9         & Error in $\langle K\rangle$ \\
10         & Number of J-band measurements \\
11         & Number of H-band measurements \\
12         & Number of K-band measurements \\
13         & Mean $\chi$ value from {\sc daophot} \\
14         & Mean sharpness value from {\sc daophot} \\
15         & Variability index $J$ \\
16         & Kurtosis index $K$ \\
17         & Variability index $L$ \\
18         & Estimated K-band amplitude \\
\multicolumn{2}{l}{\it Part II: multi-epoch data (356,303 lines)} \\
 1         & Star number \\
 2         & Epoch (HJD--2,450,000) \\
 3         & Filter (J, H or K) \\
 4         & Magnitude \\
 5         & Error in magnitude \\
 6         & $\chi$ value from {\sc daophot} \\
 7         & Sharpness value from {\sc daophot} \\
\hline
\end{tabular}
\end{table}

The photometric catalogue including all variable and non-variable stars is
made publicly available at the Centre de Donn\'ees astronomiques de Strasbourg
(CDS). The content is described in Table 3. It is composed of two parts, part
I comprising the mean properties of the stars and part II tabulating all the
photometry (for the benefit of generating lightcurves, for instance).

The astrometric accuracy of the catalogue is $\approx0.2^{\prime\prime}$
r.m.s., tied to the 2MASS system. This accuracy was found to be consistent
with the results from our cross-correlations with three other optical and IR
catalogues (cf.\ Section 6.2).

\section{Discussion}

\subsection{The near-IR variable star population}

%
%
\begin{figure}
\centerline{\psfig{figure=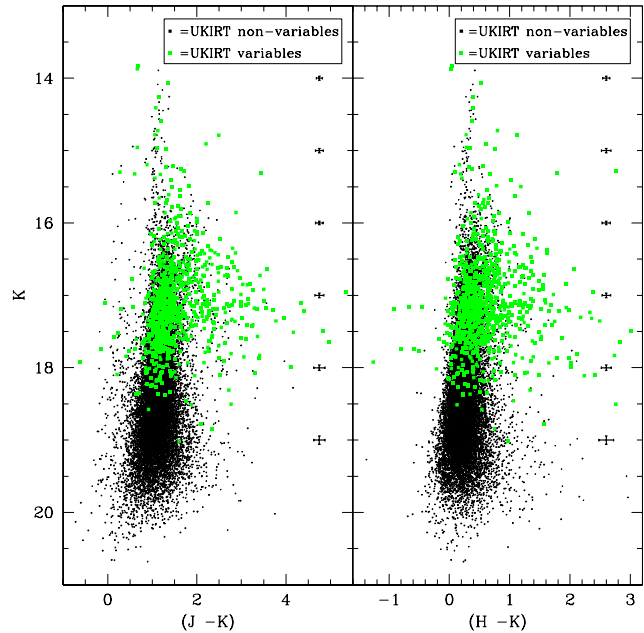,width=84mm}}
\caption[]{Near-IR colour--magnitude diagrams showing the UKIRT variable stars
in green. Average errorbars are plotted for 1-mag intervals in the K-band.}
\end{figure}

Fig.\ 13 presents near-IR colour--magnitude diagrams for our field in M\,33,
with highlighted in green the variable stars we identified. Large-amplitude
variable stars are mainly found between $K\sim16$--18 mag, and are
conspicuously absent among fainter stars. Some very bright variable stars are
found, too. Striking is the predominance of variable stars among the redder
stars that lie to the right of the vertical sequence of the bulk of the stars.

%
%
\begin{figure}
\centerline{\psfig{figure=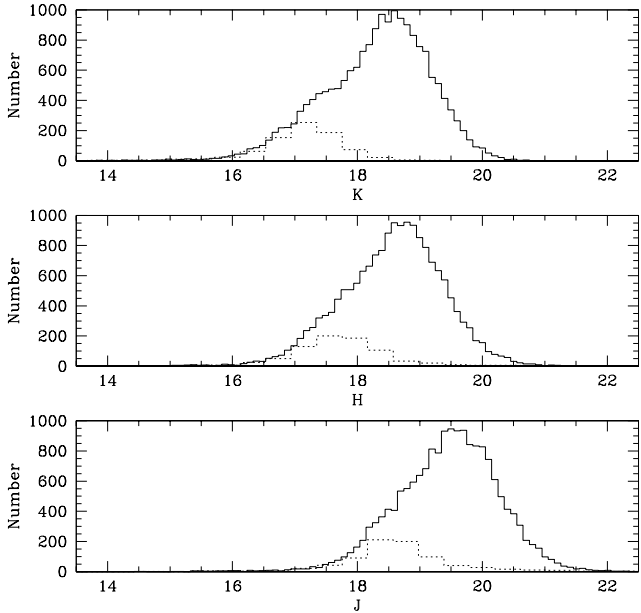,width=84mm}}
\caption[]{Distribution of all UKIRT sources (solid) and UKIRT variable stars
(dotted), as a function of near-IR brightness.}
\end{figure}

%
%
\begin{figure}
\centerline{\psfig{figure=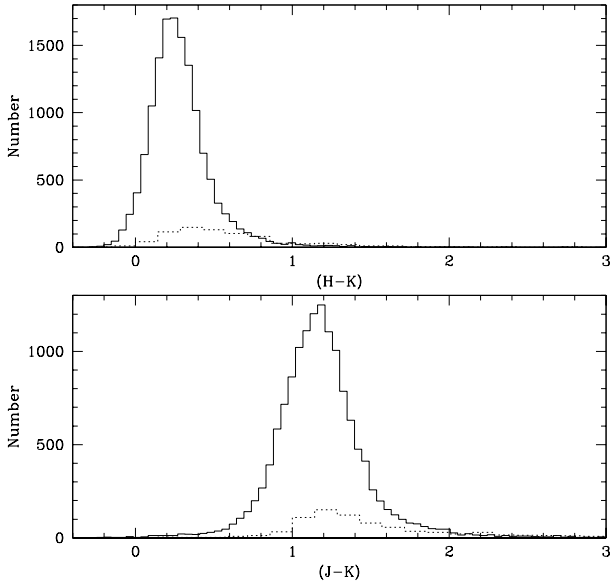,width=80mm}}
\caption[]{Distribution of all UKIRT sources with $K<19$ mag (solid) and UKIRT
variable stars (dotted), as a function of near-IR colour.}
\end{figure}

Better quantified, albeit degenerate, are histograms of the distributions over
brightness (Fig.\ 14) and colour (Fig.\ 15). Between $K\sim16$--17 mag, and
likewise for $(H-K)>0.7$ mag, the number of variable stars per histogram bin
reaches the same value as the number of all UKIRT sources per bin -- note,
however, that the bins in the former sample a three times larger range in
brightness or colour, so the true fraction of variables among those stars is
closer to a third. Still, at $(J-K)>2$ mag or $(H-K)>1$ mag (and with $K<19$
mag) almost all stars are variable.

Between $K\sim17$--18 mag the fraction of variable stars drops to only a few
per cent; the frequency of stars increases but the frequency of variable stars
decreases. This is due mainly to a combination of two factors. Firstly, lower
on the AGB there is a greater contribution from stars that will still evolve
to higher luminosity and lower temperature (thus exacerbating the brightness
increase in the K-band) before they develop large-amplitude variability. That
is precisely the reason why our project aims to use the variable stars as
tracers of the distribution of stars over birth-mass. Secondly, the relation
between birth-mass and K-band brightness flattens dramatically for low-mass
(relatively faint) AGB stars. This is explored in much more detail in Paper
II.

%
%
\begin{figure}
\centerline{\psfig{figure=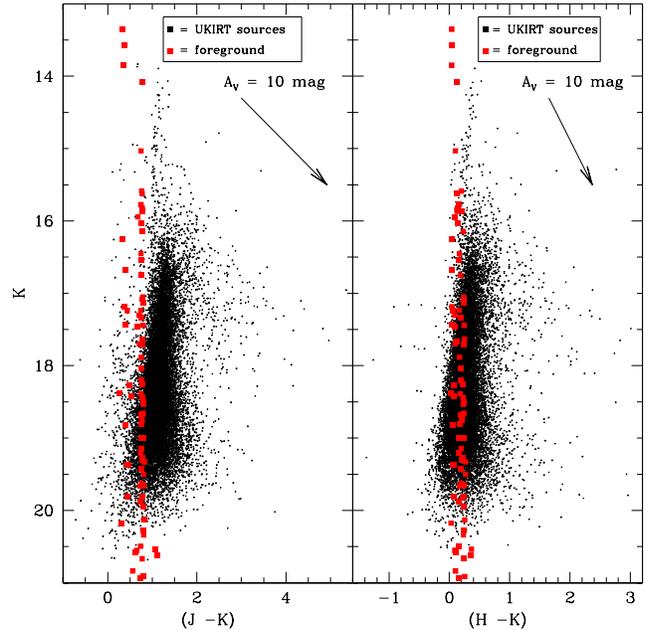,width=84mm}}
\caption[]{Estimated contamination by foreground stars (in red), from a
simulation with {\sc trilegal} (Girardi et al.\ 2005).}
\end{figure}

To assess the level of contamination by foreground stars, we performed a
simulation with the {\sc trilegal} tool (Girardi et al.\ 2005). Adopting
default parameters for the structure of the Galaxy, in a 0.005 deg$^2$ field
in the direction ($l=133.61^\circ$, $b=-31.33^\circ$) only a small number of
foreground stars are expected (Fig.\ 16). These do not generally cause
significant problems -- they all have fairly neutral colours -- except that
very bright stars at $K<14$ mag are likely to be foreground stars and not
stars within M\,33. Also indicated in Fig.\ 16 are the reddening vectors
equivalent to a visual extinction of $A_{\rm V}=10$ mag, adopting extinction
coefficients of $A_{\rm K}=0.12 A_{\rm V}$, $A_{\rm H}=1.6 A_{\rm K}$ and
$A_{\rm J}=2.6 A_{\rm K}$ (cf.\ Becklin et al.\ 1978; Savage \& Mathis 1979;
Rieke, Rieke \& Paul 1989).

%
%
\begin{figure}
\centerline{\psfig{figure=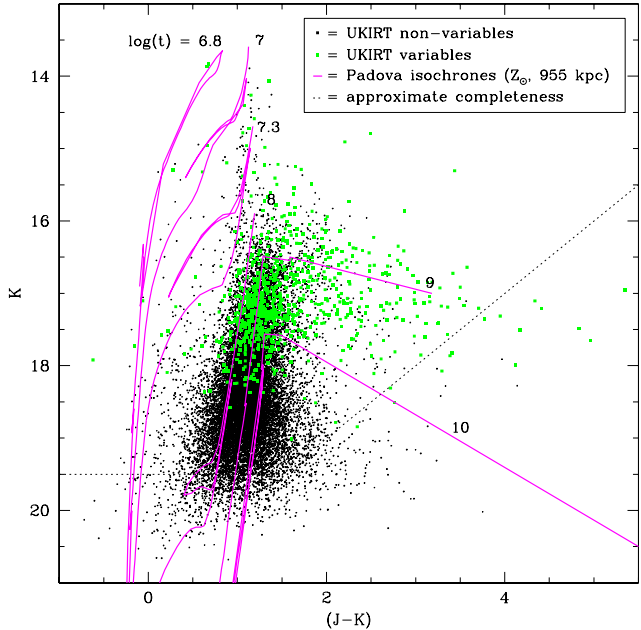,width=84mm}}
\caption[]{Colour--magnitude diagram of $(J-K)$, with UKIRT variable stars
highlighted in green. Overplotted are isochrones from Marigo et al.\ (2008)
for solar metallicity and a distance modulus of 24.9 mag.}
\end{figure}

%
%
\begin{figure}
\centerline{\psfig{figure=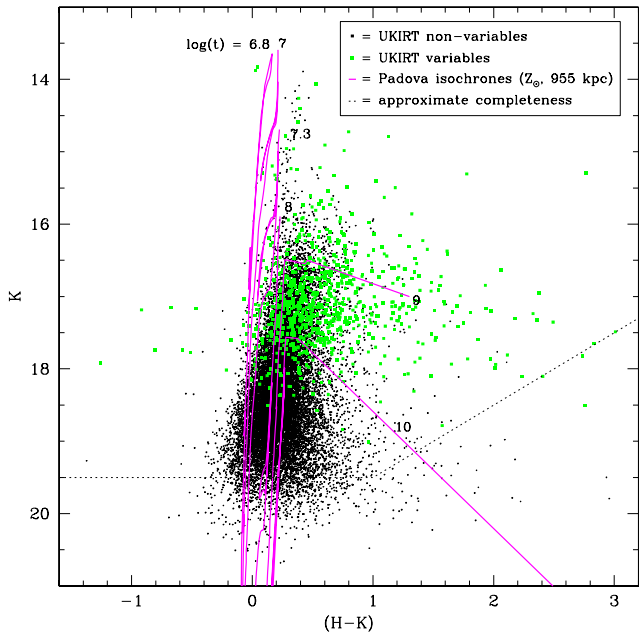,width=84mm}}
\caption[]{Colour--magnitude diagram of $(H-K)$, with UKIRT variable stars
highlighted in green. Overplotted are isochrones from Marigo et al.\ (2008)
for solar metallicity and a distance modulus of $\mu=24.9$ mag.}
\end{figure}

We characterise the stellar population in the central regions of M\,33 further
by confrontation of the colour--magnitude diagrams with isochrones calculated
by Marigo et al.\ (2008) (Figs.\ 17 and 18). The isochrones were calculated
for solar metallicity, which is appropriate for the central disc population in
M\,33 (cf.\ Magrini et al.\ 2007; Rosolowsky \& Simon 2008); a check was made
for a metallicity half that of the Sun and no appreciable difference was found
in the red (super-)giant branches. The adopted distance modulus of $\mu=24.9$
mag appears to be appropriate as it fits well with the magnitude at which the
variable stars become abundant and the K-band luminosity function shows a
break in the slope (Fig.\ 14), at the location of the RGB tip, as well as to
the maximum extent in luminosity of the RSGs. Indeed, RSGs are found all the
way until the earliest age (highest birth mass) at which they are expected,
$t\approx10$ Myr.

The isochrones of Marigo et al.\ (2008) are the most realistic models easily
available for comparison with dust-producing stellar populations, as they
include predictions for the onset of large-amplitude pulsation, the associated
mass-loss rate and dust-formation rate, and the resulting changes in the
spectral energy distribution (SED) as the dust redistributes optical light
over the IR domain. This causes the drastic excursions of the isochrones
towards red colours as soon as the dust envelope becomes optically thick at
near-IR wavelengths; hence it is accompanied by a drop in K-band brightness as
the peak of the SED shift towards (even) longer wavelengths. The 1-Gyr
isochrones show consistency with the red branch of UKIRT variables. The 10-Gyr
isochrones seem to show a rather too rapid decline in the K band compared to
observations, but this may be due to the fact that these stars experience very
large reddening and thus become readily undetectable -- this is corroborated
by the isochrones as well as by empirical evidence of low-mass AGB stars
becoming obscured by circumstellar dust much more readily than their more
massive siblings (van Loon et al.\ 1997).

\subsection{Cross-identifications in other catalogues}

We cross-correlate our UKIRT variability search results with those from two
intensive optical monitoring campaigns (CFHT, Hartman et al.\ 2006; DIRECT,
Macri et al.\ 2001) and the mid-IR variability search performed with the {\it
Spitzer} Space Telescope (McQuinn et al.\ 2007). We also compare with the
optical catalogue of Rowe et al.\ (2005) which includes narrow-band filters
that they used to identify carbon stars. The matches were obtained by search
iterations using growing search radii, in steps of $0.1^{\prime\prime}$ out to
$1^{\prime\prime}$, on a first-encountered first-associated basis after
ordering the principal photometry in order of diminishing brightness (K-band
for the UKIRT catalogue, i-band/I-band for the optical catalogues, and
3.6-$\mu$m band for the {\it Spitzer} catalogue).

\subsubsection{CFHT optical variability survey}

%
%
\begin{figure}
\centerline{\psfig{figure=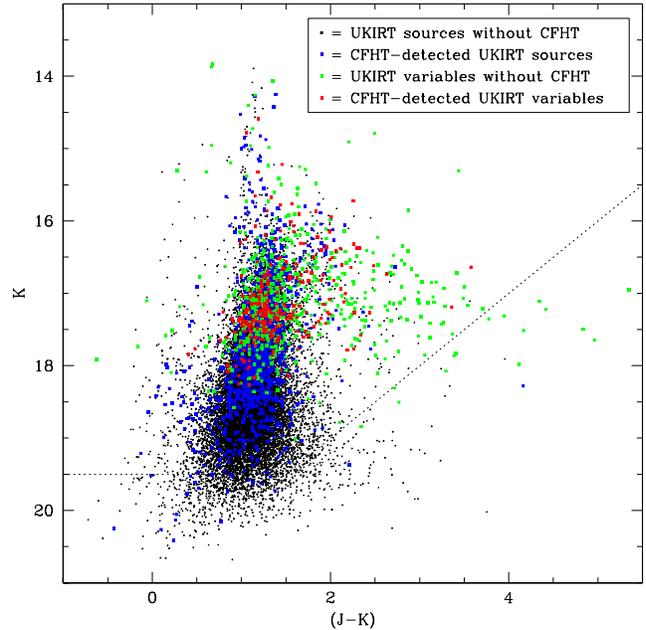,width=84mm}}
\caption[]{Near-IR colour--magnitude diagram showing the stars from the UKIRT
survey that were and were not identified as variable stars in the CFHT optical
variability survey (Hartman et al.\ 2006).}
\end{figure}

The Canada France Hawai'i Telescope (CFHT) optical variability survey (Hartman
et al.\ 2006) was performed during 27 nights (comprising 36 individual
measurements) between August 2003 and January 2005. The catalogue contains
variable stars only, with photometry in the Sloan g$^\prime$-, r$^{\prime}$-
and i$^\prime$-bands to a depth of approximately $i^\prime\approx24$ mag. Out
of 2 million point sources in a square-degree field, they identified $>1300$
candidate variable blue and red supergiants, $>2000$ Cepheids and $>19,000$
AGB and RGB LPVs.

Within the coverage of our UIST observations are located $\approx1737$
variable stars from the CFHT catalogue. Out of these, 1481 (85\%) were
detected in our UKIRT survey, of which 247 were found by us to be variable.
The CFHT variables that we did not identify as being variable are generally
fainter than the RGB tip (Fig.\ 19) and thus not the LPVs we aimed to find for
the purpose of our project, but they also include bright AGB stars and RSGs
whose modest amplitudes (especially at IR wavelengths) may have led them to
escape from our UKIRT variability search. Clearly, the dusty (reddened) AGB
variables detected in our UKIRT survey were generally not detected in the CFHT
survey.

\subsubsection{DIRECT optical variability survey}

%
%
\begin{figure}
\centerline{\psfig{figure=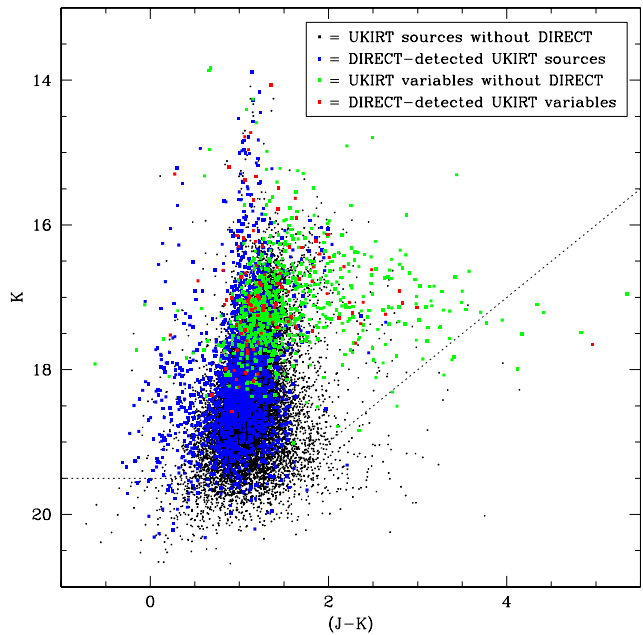,width=84mm}}
\caption[]{Near-IR colour--magnitude diagram showing the stars from the UKIRT
survey that were and were not detected in the DIRECT optical variability
survey (Macri et al.\ 2001).}
\end{figure}

The DIRECT optical variability survey (Macri et al.\ 2001) aimed to determine
direct distances to M\,31 and M\,33 using detached eclipsing binaries and
Cepheid variables. The catalogue is based on observations performed between
September 1996 and October 1997, during 95 nights on the F.\ L.\ Whipple
Observatory 1.2-m telescope and 36 nights on the Michigan--Dartmouth--MIT
1.3-m telescope. The catalogue contains Johnson B- and V-, and Cousins I-band
photometry for all stars with $14.4<V<23.6$ mag, and lists the V-band $J$
variability index (cf.\ Section 4).

Within the coverage of our UIST observations are located $\approx2644$ stars
from the DIRECT catalogue (among which 113 have $J>0.75$, which the DIRECT
survey team considered as the threshold for variability). Out of these, 2018
(76\%) were detected in our UKIRT survey, of which 106 were found by us to be
variable. Remarkably, most of the UKIRT variables (87\%) were missed by the
DIRECT survey (Fig.\ 20). This is due in part because the dusty variables are
very faint in the optical, but surprisingly most of the not-so-dusty AGB
variables are also missing from the DIRECT survey. The DIRECT survey seems to
have been less successful in finding LPVs than the CFHT survey.

\subsubsection{Carbon star survey}

Rowe et al.\ (2005) used the CFHT and a four-filter system in 1999 and 2000 to
cover most of M\,33 including the central region. The filter system was
designed to identify carbon stars on the basis of their cyanide (CN)
absorption as opposed to other red giants that display titanium-oxide (TiO)
absorption, using narrow-band filters centred on 8120 and 7777 \AA,
respectively. They added broad-band Mould V- and I-band filters to aid in
selecting cool stars. Carbon stars in their scheme have [CN]$-$[TiO] $>0.3$
and $V-I>1.8$ mag, and M-type stars have [CN]$-$[TiO] $<-0.2$ mag at the same
V--I criterion (they only considered stars for this purpose that had errors on
these colours of $<0.05$ mag).

The region covered by our UKIRT survey contains $\sim$100,000 stars from the
Rowe et al.\ catalogue; all but 837 of our objects were identified among
these. The problem with such density of optical sources is that chance
co-incidences are common. When a pre-selection is made of stars that are
probably carbon stars or M-type stars, which are likely to have been detected
at near-IR wavelengths, then the number of stars to correlate with is
drastically reduced. The average offsets between these and their identified
UKIRT counterparts are only $0.02^{\prime\prime}$ in both RA and Dec.

%
%
\begin{figure}
\centerline{\psfig{figure=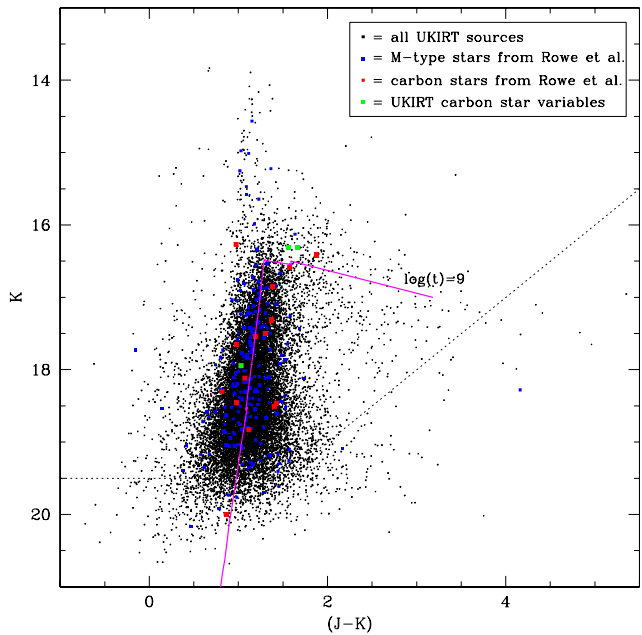,width=84mm}}
\caption[]{Near-IR colour--magnitude diagram showing the M-type and carbon
stars from the Rowe et al.\ (2005) survey that were detected in our UKIRT
survey. A 1-Gyr isochrone from Marigo et al.\ (2008) is shown for comparison.}
\end{figure}

The M-type stars follow the main giant branch (Fig.\ 21), avoiding the blue
edge which is expected to be occupied by K-type giants without strong
molecular absorption. The carbon stars are consistent with ages of $t\sim1$
Gyr or a little younger, i.e.\ birth masses around 2--3 M$_\odot$. Two of the
three carbon stars that are identified as UKIRT variables show signs of
reddening presumably due to circumstellar dust; these are also among the
brightest carbon stars and thus represent the termination point in their
evolution. The faintest carbon stars are fainter than the tip of the RGB;
while unexpected for carbon stars formed by thermal pulses on the AGB, faint
carbon stars are known in other, generally metal-poor, populations, e.g., in
the Sagittarius dwarf iregular galaxy (Gullieuszik et al.\ 2007) or in the
Galactic globular cluster $\omega$\,Centauri (van Loon et al.\ 2007).

%
%
\begin{figure}
\centerline{\psfig{figure=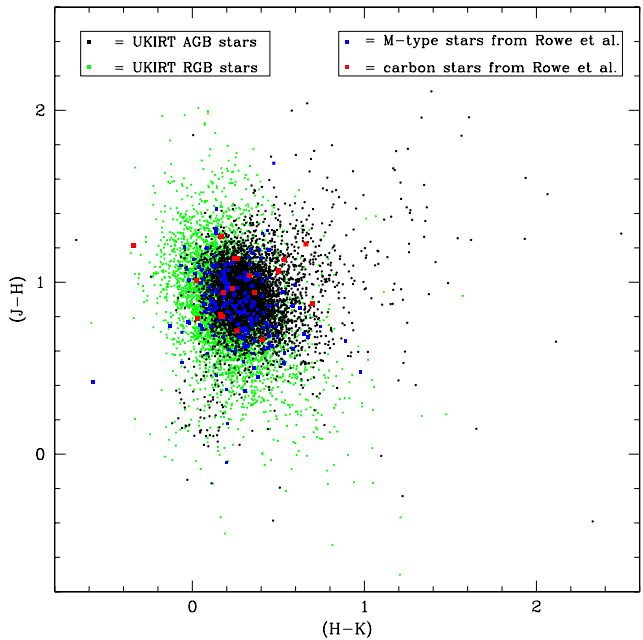,width=84mm}}
\caption[]{Near-IR colour--colour diagram showing the M-type and carbon
stars from the Rowe et al.\ (2005) survey that were detected in our UKIRT
survey. Stars with $18<K<19.5$ mag and errors on the colours $<0.25$ mag are
labelled as RGB stars, whilst stars with $16<K<18$ mag and errors on the
colours $<0.25$ mag are labelled as AGB stars}
\end{figure}

A colour-colour representation is shown in Fig.\ 22, where a crude distinction
is made between RGB and AGB stars. Again, the reddened stars are AGB stars;
RGB stars displaying a red J--K colour usually do not display a red H--K
colour (or {\it vice versa}), suggesting that they are not (much) affected by
reddening but that photometric uncertainties become larger. The M-type and
carbon stars from Rowe et al.\ (2005) are generally constrained to the AGB,
with a few carbon stars showing signs of reddening.

Though not many stars in Rowe et al.\ have errors on their colours of $<0.05$
mag, and the suspicion is that many more of the UKIRT sources are carbon stars
or M-type stars, carbon stars do not seem to dominate the stellar population
in the central square kpc of M\,33. This onfirms the trend of the ratio of
carbon to M-type stars found by Rowe et al.\ (2005) to drop from typically
$\approx0.5$ to $<0.2$ in the central few arcminutes, and it is corroborating
evidence for a predominantly solar-metallicity population as carbon stars are
more common among populations with sub-solar metallicity (Groenewegen 1999).

\subsubsection{{\it Spitzer} mid-IR variability survey}

McQuinn et al.\ (2007) used five epochs of {\it Spitzer} Space Telescope
observations of M\,33 obtained in the 3.6-, 4.5- and 8-$\mu$m bands, to
identify variable stars in a manner which is broadly similar to our approach
with UKIRT.

%
%
\begin{figure}
\centerline{\psfig{figure=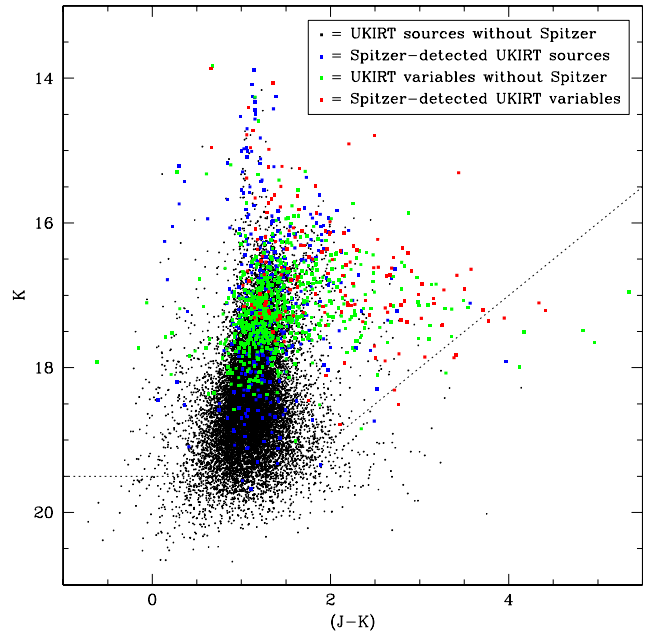,width=84mm}}
\caption[]{Near-IR colour--magnitude diagram showing the stars from the UKIRT
survey that were and were not detected in the {\it Spitzer} variability survey
(McQuinn et al.\ 2007).}
\end{figure}

In the central regions of M\,33, crowding and strong, complex diffuse emission
severely limited the capability of the {\it Spitzer} survey. It yielded 784
sources in roughly the same area covered by our UIST data. Among these, 557
are in our photometric catalogue, most of them are brighter than the RGB tip.
Of the 227 {\it Spitzer} sources that we did not recover, 13 had been rejected
from our catalogue because of too high $\chi$ values -- these were at the
edges of frames and only covered in part. Hence the recovery rate is well
above 70\%. The success rate of the {\it Spitzer} survey in detecting stars
from our UKIRT survey is particularly good for the RSGs, the brightest AGB
stars, and the dusty AGB variables (Fig.\ 23); AGB stars not in the {\it
Spitzer} catalogue are generally situated closer to the centre of M\,33 where
crowding increases. Indeed, blending is quite common, with the {\it Spitzer}
source being centred in between the position of two similarly-bright stars.

%
%
\begin{figure}
\centerline{\psfig{figure=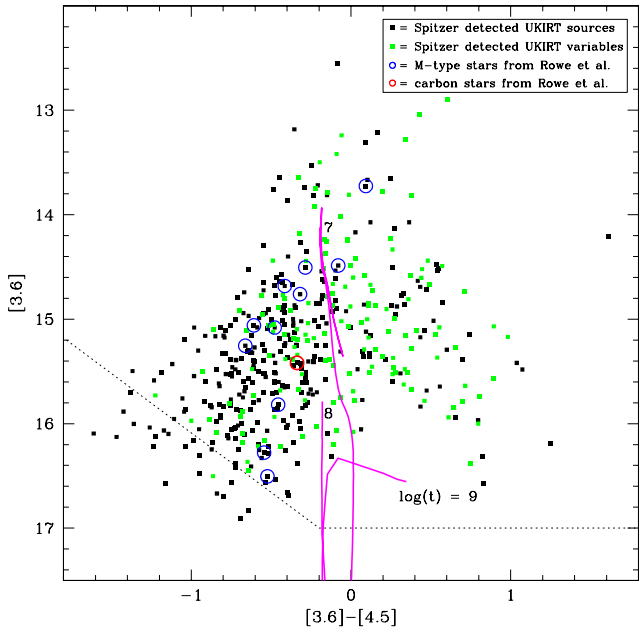,width=84mm}}
\caption[]{Mid-IR colour--magnitude diagram from {\it Spitzer} photometry of
UKIRT sources, with UKIRT variables highlighted in green and M-type and carbon
stars from Rowe et al.\ (2005) circled in blue and red, respectively.
Isochrones from Marigo et al.\ (2008) for 10 Myr, 100 Myr and 1 Gyr are drawn.
The dotted line indicates the approximate detection limit of the {\it Spitzer}
survey.}
\end{figure}

The mid-IR colour--magnitude diagram (Fig.\ 24) shows a well-populated,
skewed sequence off from which a branch extends towards redder colour and
fainter 3.6-$\mu$m brightness. The latter is comprised of dust-enshrouded
objects, among which a relatively greater proportion are UKIRT variables.
UKIRT variables are also abound among the brightest 3.6-$\mu$m sources in the
diagram; these are massive AGB stars and RSGs. The confirmed M-type stars form
a sequence of increasing 3.6-$\mu$m brightness, and the only confirmed carbon
star that is detected by {\it Spitzer} (and us) sits roughly half-way and
marginally to the red of that sequence. None of these are on the red branch of
dust-enshrouded objects, probably because the optical photometry was not deep
enough for them to have been detected or not accurate enough for them to have
been classified.

The isochrones confirm the picture just outlined in a qualitative manner, but
there is a surprisingly large discrepancy between the {\it Spitzer} photometry
and the isochrones. The {\it Spitzer} photometry appears too blue at faint
magnitudes, by at least several tenths of a magnitude -- we consider it
unlikely that the colours are really as negative as $([3.6]-[4.5])\approx-0.7$
mag (a factor two depression of the 4.5-$\mu$m brightness). For the isochrones
to match the 3.6-$\mu$m brightness, in agreement with the way in which they
match (well) the near-IR colour--magnitude diagrams, the isochrones would need
to be shifted to higher 3.6-$\mu$m brightness levels by at least a magnitude
or the data need to be diminished by that amount. We find either solution
unsatisfactory, and suspect that both the {\it Spitzer} photometry and
isochrones may need adjustments. We postpone a thorough investigation to Paper
III in which we model the SEDs.

\section{Conclusions}

UKIRT was used to monitor the central $4^\prime\times4^\prime$ (square kpc) of
Local Group spiral galaxy M\,33 (Triangulum) in the K-band filter with
additional observations in the J- and H-band filters.

As a result, a photometric catalogue was compiled of 18,398 stars among which
812 were identified as exhibiting large-amplitude variability. Inspection of
the lightcurves and locations on the colour--magnitude diagram with respect to
theoretical models of stellar evolution leads us to conclude that most of
these variable stars are AGB stars or RSGs, and that most of the very dusty
stars -- that are heavily reddened even at IR wavelengths -- are variable.

Our catalogue is cross-correlated with previous optical monitoring surveys, an
optical carbon star survey, and the {\it Spitzer} mid-IR survey. The UKIRT
catalogue is vastly more complete for the dusty variables than the optical
surveys, but also much more complete than the {\it Spitzer} survey. Our
catalogue is made public at CDS.

In the next papers in this series, our catalogue will be used to describe the
star formation history and the dust production in the central regions of
M\,33.

\section*{Acknowledgments}

We thank the staff at UKIRT for their excellent support, and the visiting
astronomers who executed this programme in queue observing mode. AJ is
grateful to Peter Stetson for sharing his photometry routines and for valuable
advice. Joana Oliveira helped us understand the photometric errors. Jason Rowe
is thanked for sending us his full catalogue. We are indebted to Albert
Zijlstra, Michael Feast and Patricia Whitelock for advice in the initial
stages of the project. We also thank the anonymous referee for her/his
suggestions that helped improve the presentation of the manuscript. JvL wishes
to thank the Iranian astronomers and students for their kindness during his
stay at the Institute of Physics and Mathematics in Tehran and during his
visits to beautiful Shiraz, Esfahan and Yazd. This project was made possible
through financial support by The Leverhulme Trust under grant No.\
RF/4/RFG/2007/0297.


\label{lastpage}


\begin{thebibliography}{}
\bibitem[Becklin et al.(1978)]{Becklin78}
Becklin E.\ E., Neugebauer G., Willner S.\ P., Matthews K., 1978, ApJ, 220,
831
\bibitem[Bonanos et al.(2006)]{Bonanos06}
Bonanos A.\ Z.\ et al., 2006, ApJ, 652, 313
\bibitem[Bowen (1988)]{Bowen88}
Bowen G.\ H., 1988, ApJ, 329, 299
\bibitem[Bowen \& Willson (1991)]{Bowen91}
Bowen G.\ H., Willson L.\ A., 1991, ApJ, 375, L53
\bibitem[Carpenter (2001)]{Carpenter01}
Carpenter J.\ M., 2001, AJ, 121, 2851
\bibitem[Freedman, Wilson \& Madore (1991)]{Freedman91}
Freedman W.\ L., Wilson C.\ D., Madore B.\ F., 1991, ApJ, 372, 455
\bibitem[Galleti, Bellazzini \& Ferraro (2004)]{Galleti04}
Galleti S., Bellazzini M., Ferraro F.\ R., 2004, A\&A, 423, 925
\bibitem[Girardi et al.(2005)]{Girardi05}
Girardi L., Groenewegen M.\ A.\ T., Hatziminaoglou E., da Costa L., 2005,
A\&A, 436, 895
\bibitem[Groenewegen (1999)]{Groenewegen99}
Groenewegen M.\ A.\ T., in: Asymptotic Giant Branch Stars, eds.\ T.\ Le
Bertre, A.\ L\`ebre and C.\ Waelkens, IAU Symposium 191, p535
\bibitem[Gullieuszik et al.(2007)]{Gullieuszik07}
Gullieuszik M., Rejkuba M., Cioni M.\ R., Habing H.\ J., Held E.\ V., 2007,
A\&A, 475, 467
\bibitem[Hartman et al.(2006)]{Hartman06}
Hartman J.\ D., Bersier D., Stanek K.\ Z., Beaulieu J.-P., Ka{\l}u$\dot{\rm
z}$ny J., Marquette J.-B., Stetson P.\ B., Schwarzenberg-Czerny A., 2006,
MNRAS, 371, 1405
\bibitem[Kim et al.(2002)]{Kim02}
Kim M., Kim E., Lee M.\ G., Sarajedini A., Geisler D., 2002, AJ, 123, 244
\bibitem[Krisciunas et al.(1987)]{Krisciunas87}
Krisciunas K.\ et al., 1987, PASP, 99, 887
\bibitem[Macri et al.(2001)]{Macri01}
Macri L.\ M., Stanek K.\ Z., Sasselov D.\ D., Krockenberger M.,
Ka{\l}u$\dot{\rm z}$ny J., 2001, AJ, 121, 861
\bibitem[Levesque (2010)]{Levesque10}
Levesque E.\ M., 2010, in: Hot and Cool -- Bridging Gaps in Massive Star
Evolution, eds.\ C.\ Leitherer, P.\ Bennett, P.\ Morris and J.\ Th.\ van Loon,
ASPC (San Francisco: ASP), 425, p103
\bibitem[Levesque et al.(2005)]{Levesque05}
Levesque E.\ M., Massey P., Olsen K.\ A.\ G., Plez B., Josselin E., Maeder A.,
Meynet G., 2005, ApJ, 628, 973
\bibitem[Magrini et al.(2007)]{Magrini07}
Magrini L., V\'{\i}lchez J.\ M., Mampaso A., Corradi R.\ L.\ M., Leisy P.,
2007, A\&A, 470, 865
\bibitem[Marigo et al.(2008)]{Marigo08}
Marigo P., Girardi L., Bressan A., Groenewegen M.\ A.\ T., Silva L., Granato
G.\ L., 2008, A\&A, 482, 883
\bibitem[McConnachie et al.(2004)]{Mcconnachie04}
McConnachie A.\ W., Irwin M.\ J., Ferguson A.\ M.\ N., Ibata R.\ A., Lewis G.\
F., Tanvir N., 2004, MNRAS, 350, 243
\bibitem[McQuinn et al.(2007)]{Mcquinn07}
McQuinn K.\ B.\ W.\ et al., 2007, ApJ, 664, 850
\bibitem[Mochejska et al.(2001a)]{Mochejska01a}
Mochejska B.\ J., Ka{\l}u$\dot{\rm z}$ny J., Stanek K.\ Z., Sasselov D.\ D.,
Szentgyorgyi A.\ H., 2001a, AJ, 121, 2032
\bibitem[Mochejska et al.(2001b)]{Mochejska01b}
Mochejska B.\ J., Ka{\l}u$\dot{\rm z}$ny J., Stanek K.\ Z., Sasselov D.\ D.,
Szentgyorgyi A.\ H., 2001b, AJ, 122, 2477
\bibitem[Pierce, Jurcevi\' \& Crabtree (2000)]{Pierce00}
Pierce M.\ J., Jurcevi\'c J.\ S., Crabtree D., 2000, MNRAS, 313, 271
\bibitem[Rieke, Rieke \& Paul (1989)]{Rieke89}
Rieke G.\ H., Rieke M.\ J., Paul A.\ E., 1989, ApJ, 336, 752
\bibitem[Rizzi et al.(2007)]{Rizzi07}
Rizzi L., Tully R.\ B., Makarov D., Makarova L., Dolphin A.\ E., Sakai S.,
Shaya E.\ J., 2007, ApJ, 661, 815
\bibitem[Rosolowsky \& Simon (2008)]{Rosolowsky08}
Rosolowsky E., Simon J.\ D., 2008, ApJ, 675, 1213
\bibitem[Rowe et al.(2005)]{Rowe05}
Rowe J.\ F., Richer H.\ B., Brewer J.\ P., Crabtree D.\ R., 2005, AJ, 129, 729
\bibitem[Sarajedini et al.(2000)]{Sarajedini00}
Sarajedini A., Geisler D., Schommer R., Harding P., 2000, AJ, 120, 2437
\bibitem[Sarajedini et al.(2006)]{Sarajedini06}
Sarajedini A., Barker M.\ K., Geisler D., Harding P., Schommer R., 2006, AJ,
132, 1361
\bibitem[Savage \& Mathis (1979)]{Savage79}
Savage B.\ D., Mathis J.\ S., 1979, ARA\&A, 17, 73
\bibitem[Scowcroft et al.(2009)]{Scowcroft09}
Scowcroft V., Bersier D., Moul J.\ R., Wood P.\ R., 2009, MNRAS, 396, 1287
\bibitem[Stetson (1987)]{Stetson87}
Stetson P.\ B., 1987, PASP, 99, 191
\bibitem[Stetson (1990)]{Stetson90}
Stetson P.\ B., 1990, PASP, 102, 932
\bibitem[Stetson (1993)]{Stetson93}
Stetson P.\ B., 1993, in: Stellar Photometry -- Current Techniques and Future
Developments, eds.\ C.\ J.\ Butler and I.\ Elliott, IAU Coll.\ Ser.\ 136
(Cambridge: Cambridge University Press), p291
\bibitem[Stetson (1994)]{Stetson94}
Stetson P.\ B., 1994, PASP, 106, 250
\bibitem[Stetson (1996)]{Stetson96}
Stetson P.\ B., 1996, PASP, 108, 851
\bibitem[Tiede, Sarajedini \& Barker (2004)]{Tiede04}
Tiede G.\ P., Sarajedini A., Barker M.\ K., 2004, AJ, 128, 224
\bibitem[U et al.(2009)]{U09}
U V., Urbaneja M.\ A., Kudritzki R.-P., Jacobs B.\ A., Bresolin F., Przybilla
N., 2009, ApJ, 704, 1120
\bibitem[van Loon (2010)]{vanLoon10}
van Loon J.\ Th., 2010, in: Hot and Cool -- Bridging Gaps in Massive Star
Evolution, eds.\ C.\ Leitherer, P.\ Bennett, P.\ Morris and J.\ Th.\ van Loon,
ASPC (San Francisco: ASP), 425, p279
\bibitem[van Loon et al.(1997)]{vanLoon97}
van Loon J.\ Th., Zijlstra A.\ A., Whitelock P.\ A., Waters L.\ B.\ F.\ M.,
Loup C., Trams N.\ R., 1997, A\&A, 325, 585
\bibitem[van Loon et al.(1999)]{vanLoon99}
van Loon J.\ Th., Groenewegen M.\ A.\ T., de Koter A., Trams N.\ R., Waters
L.\ B.\ F.\ M., Zijlstra A.\ A., Whitelock P.\ A., Loup C., 1999, A\&A, 351,
559
\bibitem[van Loon et al.(2005)]{vanLoon05}
van Loon J.\ Th., Cioni M.-R.\ L., Zijlstra A.\ A., Loup C., 2005, A\&A, 438,
273
\bibitem[van Loon et al.(2007)]{vanLoon07}
van Loon J.\ Th., van Leeuwen F., Smalley B., Smith A.\ W., Lyons N.\ A.,
McDonald I., Boyer M.\ L., 2007, MNRAS, 382, 1353
\bibitem[van Loon et al.(2008)]{vanLoon08}
van Loon J.\ Th., Cohen M., Oliveira J.\ M., Matsuura M., McDonald I., Sloan
G.\ C., Wood P.\ R., Zijlstra A.\ A., 2008, A\&A, 487, 1055
\bibitem[Vassiliadis \& Wood (1993)]{Vassiliadis93}
Vassiliadis E., Wood P.\ R., 1993, ApJ, 413, 641
\bibitem[Whitelock et al.(2003)]{Whitelock03}
Whitelock P.\ A., Feast M.\ W., van Loon J.\ Th., Zijlstra A.\ A., 2003,
MNRAS, 342, 86
\bibitem[Wood (1998)]{Wood98}
Wood P.\ R., 1998, A\&A, 338, 592
\bibitem[Wood (1999)]{Wood99}
Wood P.\ R., in: Asymptotic Giant Branch Stars, eds.\ T.\ Le Bertre, A.\
L\`ebre and C.\ Waelkens, IAU Symposium 191, p151
\bibitem[Wood et al.(1992)]{Wood92}
Wood P.\ R., Whiteoak J.\ B., Hughes S.\ M.\ G., Bessell M.\ S., Gardner F.\
F., Hyland A.\ R., 1992, ApJ, 397, 552
\end{thebibliography}
\end{document}